\documentclass[lettersize,journal]{IEEEtran}

\usepackage{graphicx} 
\usepackage{float} 
\usepackage{subfigure} 
\usepackage{diagbox}
\usepackage{algorithm}
\usepackage{algorithmic}
\usepackage{multirow}
\usepackage{makecell}
\usepackage{amsmath}
\usepackage{caption}
\usepackage{cite}
\usepackage{color}
\usepackage{url,array}
\usepackage{amssymb}
\usepackage{hyperref}
\usepackage{booktabs}
\usepackage{amsmath}
\usepackage{hyperref}
\usepackage{bbding}
\usepackage{algorithm}
\usepackage{algorithmic}
\newcommand{\PreserveBackslash}[1]{\let\temp=\\#1\let\\=\temp}
\newcolumntype{C}[1]{>{\PreserveBackslash\centering}p{#1}}
\newcolumntype{R}[1]{>{\PreserveBackslash\raggedleft}p{#1}}
\newcolumntype{L}[1]{>{\PreserveBackslash\raggedright}p{#1}}
\begin{document}

\title{Distinctive and Natural Speaker Anonymization via Singular Value Transformation-assisted Matrix}

\author{Jixun Yao, Qing Wang, Pengcheng Guo, Ziqian Ning,~Lei~Xie,~\IEEEmembership{Senior member,~IEEE,}
\thanks{
Manuscript received 15 August 2023; revised 17 February 2024; accepted 15 May 2024. The associate editor coordinating the review of this manuscript and approving it for publication was Sadjadi, Omid. (\textit{Corresponding author: Lei Xie.})

The authors are with the Audio, Speech and Langauge Processing Group (ASLP@NPU), School of Computer Science, Northwestern Polytechnical
University, Xi’an 710129, China (e-mail: yaojx@mail.nwpu.edu.cn;  qingwang@nwpu-aslp.org; guopengcheng1220@gmail.com; ningziqian@
mail.nwpu.edu.cn; lxie@nwpu.edu.cn).
}
}


%
%

\markboth{Journal of \LaTeX\ Class Files,~Vol.~14, No.~8, August~2015}%
{Shell \MakeLowercase{\textit{et al.}}: Bare Demo of IEEEtran.cls for IEEE Journals}
%



\maketitle
\begin{abstract}
Speaker anonymization is an effective privacy protection solution that aims to conceal the speaker's identity while preserving the naturalness and distinctiveness of the original speech. Mainstream approaches use an utterance-level vector from a pre-trained automatic speaker verification (ASV) model to represent speaker identity, which is then averaged or modified for anonymization.
However, these systems suffer from deterioration in the \textit{naturalness} of anonymized speech, degradation in speaker \textit{distinctiveness}, and severe privacy leakage against powerful attackers. 
To address these issues and especially generate more natural and distinctive anonymized speech, we propose a novel speaker anonymization approach that models a matrix related to speaker identity and transforms it into an anonymized singular value transformation-assisted matrix to conceal the original speaker identity. Our approach extracts frame-level speaker vectors from a pre-trained ASV model and employs an attention mechanism to create a speaker-score matrix and speaker-related tokens. Notably, the speaker-score matrix acts as the weight for the corresponding speaker-related token, representing the speaker's identity. The singular value transformation-assisted matrix is generated by recomposing the decomposed orthonormal eigenvectors matrix and non-linear transformed singular through Singular Value Decomposition (SVD). This process prevents the degradation of speaker distinctiveness caused by introducing other speakers' identity information. By multiplying the singular value transformation-assisted matrix and speaker-related tokens, we generate the anonymized speaker identity representation, thereby producing anonymized speech that is both natural and distinctive. 
Experiments on VoicePrivacy Challenge datasets demonstrate the effectiveness of our approach in protecting speaker privacy under all attack scenarios while maintaining speech naturalness and distinctiveness.
\end{abstract}

\begin{IEEEkeywords}
Speaker anonymization, privacy protection, VoicePrivacy Challenge, singular value decomposition.
\end{IEEEkeywords}

%
\IEEEpeerreviewmaketitle

\section{Introduction}
\IEEEPARstart{W}{ithin} the last decades, speech data on the Internet have proliferated exponentially due to the advent of social media. Applications such as telecommunication, voice pay, and speech assistant store speech data on a centralized server, rendering it vulnerable to theft by malicious attackers.
Speech data contain rich personal sensitive information, such as age, healthy state, religious beliefs, etc., which can be recognized by speaker~\cite{zhou2021resnext,wang2019adversarial, wang2018unsupervised}, pathological condition~\cite{dibazar2002feature,schuller2013interspeech}, or other types of speech attribute recognition systems. 
Furthermore, current neural speech synthesis techniques can generate high-fidelity and almost human-like speech, raising the possibility of re-synthesis or cloning of an individual's unique vocal identity, potentially leading to unauthorized access to voice-controlled devices.
With new regulations exemplified by the General Data Protection Regulation (GDPR) in the European Union~\cite{nautsch2019gdpr}, privacy protection of personal data has drawn more attention.
\textit{Speaker anonymization} serves as a proactive privacy protection solution that is implemented before users sharing their speech data. Its fundamental objective is to effectively cancel a speaker's identity while preserving the distinctiveness, naturalness, and intelligibility of the original speech. Intelligibility necessitates that anonymized speech retains the original linguistic content, while naturalness demands that anonymized speech preserves paralinguistic attributes unaltered, including prosody and rhythm, among others. Moreover, the distinctiveness aspect ensures that all trial utterances from a specific speaker are anonymized by the same pseudo-speaker, while trial utterances from different speakers are uttered by different pseudo-speakers~\cite{vpc2022eval}.

Until recently, speaker anonymization is still in its infancy and existing works lacked consistency in definitions and metrics~\cite{yao2023distinguishable}. To solve this problem and make anonymization systems easy to compare, VoicePrivacy Challenge (VPC)~\cite{vpc2020intro,vpc2020eval,vpc2022eval} was introduced by the speech community and held in both 2020 and 2022. VPC clearly defines the speaker anonymization task, benchmark, metrics, and datasets to drive the development and innovation of techniques dedicated to preserving the privacy of speech data. 

The VPC series has shown that the majority of current speaker anonymization works adhere to one of the established baselines, leading to their categorization into two classes: (1) signal-processing based approach~\cite{sig1, sig2, sig3} and (2) neural voice conversion based approach~\cite{nn2,nn4,nn5,nn6,nn7,nn8,nn9,nn10}. 
In the context of speaker anonymization, most current works hypothesize that the speaker's identity representation carries most of the private information and can protect privacy by modifying it accordingly. In this way, the hypothesis enables the suppression of original identity information while preserving other aspects of the original speech signal.
The evaluation conducted in the VPC indicates that neural voice conversion based anonymization approaches outperform signal-processing based approaches in terms of privacy protection~\cite{vpc2020eval,vpc2022eval}. The signal-processing-based approaches might be linked back to the original speech signal after a reasonable number of attempts because the scope of physical shifts for speech signals is limited.

A general framework for neural voice conversion based approaches typically comprises three main components. The first component is feature extraction, which involves extracting linguistic content, speaker identity, and prosody information from the input speech signal. The bottleneck feature extracted from an intermediate layer of an automatic speech recognition (ASR) model~\cite{asr1, asr2} is widely acknowledged as speaker-independent linguistic content representation. Meanwhile, a speaker-dependent vector extracted from an automatic speaker verification (ASV) model~\cite{xvector,ecapatdnn} represents speaker identity, including highly personalized information. Both of these models need to be pre-trained on extensive labeled datasets. Additionally, the prosody representation is represented through fundamental frequency (F0). The second component is the speaker anonymizer, which is responsible for transforming the original speaker vector into an anonymized vector. Typically, this involves selecting a list of candidate speaker vectors from a speaker vector pool based on certain metrics~\cite{nn4,nn9}, which are then averaged or modified to generate the anonymized speaker vector. The third component is the voice conversion model, which consists of both an acoustic model and a vocoder designed for voice conversion~\cite{fastspeech,hifigan}. This conversion model reconstructs the anonymized mel-spectrogram by leveraging the original fundamental frequency, linguistic content, and the anonymized speaker vector. And then, the vocoder synthesizes the anonymized mel-spectrogram to a high-quality anonymized speech signal.

The main challenge in achieving speaker anonymization, as highlighted by~\cite{vpc2022eval}, is selecting an appropriate speaker identity representation and effectively suppressing the original speaker information.
Previous works have proposed to use the speaker vector directly extracted from a pre-trained ASV model~\cite{nn4} or a look-up table (LUT)~\cite{nn7} as the representation of speaker identity. These representations can be further modified through techniques such as averaging~\cite{nn4,nn7}, scaling~\cite{yao2023distinguishable}, or adding adversarial noise~\cite{nn2}.
However, using the speaker vector directly extracted from a pre-trained ASV model in the voice conversion model may have limitations. Because the pre-trained ASV model and the voice conversion model are typically not jointly trained using a conversion optimization function~\cite{dgc2022}. As a result, the performance of speaker anonymization has significantly decreased against more powerful attackers. 
Moreover, the representation extracted from the ASV model is a global vector that may retain residual speaker-independent global information, such as style and prosody. When employing direct averaging or scaling of this representation, the distinctiveness and naturalness of anonymized speech experience a noticeable degradation, primarily due to the introduction of other speakers' identity information and the mismatch between the ASV model and the voice conversion model.

In this study, we propose a novel speaker anonymization approach that models speaker identity in a matrix form, named singular value transformation-assisted matrix. The singular value transformation-assisted matrix represents the weights corresponding to the speaker's identity information and can effectively conceal the speaker's identity by modifying matrix values while maintaining speech naturalness and distinctiveness. To avoid the mismatch between the pre-trained ASV model and acoustic model in the conventional approach, we extract frame-level speaker representation from a pre-trained ASV and leverage self-supervised learning (SSL) to generate the speaker-related tokens and speaker-score matrix to obtain a more appropriate utterance-level speaker identity representation. The SSL module is jointly optimized with the conversion model, which leads to an obvious improvement in the naturalness of the converted speech and can effectively capture the distinct attributes of speaker identity. On the other hand, the speaker-related token is similar to the soft speaker identity label and the speaker-score matrix is the token weight, thus we can anonymize the original speaker identity by modifying the speaker-score matrix. 
We employ Singular Value Decomposition (SVD) to decompose the speaker-score matrix, resulting in a singular matrix representing the correlation used to generate the final speaker representation weights. By transforming the singular matrix, we can modify the weights of the original speaker matrix to achieve the anonymization process. The proposed anonymization process eliminates the need for additional speaker vector pools, reducing the complexity. The SVD process prevents degradation in speaker distinctiveness caused by the introduction of other speakers' identity information.

We summarize our main contributions as follows:
\begin{itemize}
\item We propose a distinctiveness-preserved anonymization strategy by leveraging the singular value transformation-assisted matrix. This strategy eliminates the need for an additional speaker vector pool and ensures that no information from other speakers is introduced, thereby preserving the distinctiveness of the anonymized speaker.
\item To effectively capture the distinct attributes of speaker identity, we disentangle the speaker identity by extracting frame-level speaker vectors from a pre-trained ASV model and leverage the attention mechanism in SSL to obtain a speaker-score matrix and speaker-related tokens. This method allows us to anonymize the speaker representation by transforming the speaker-score matrix to conceal the speaker's identity.
\item We employ soft linguistic content and Mutual Information (MI) constraints in our speaker anonymization system to better preserve the paralinguistic information and disentangle the content, speaker identity, and prosody.
\end{itemize}


The rest of the paper is organized as follows.
In Section \uppercase\expandafter{\romannumeral2}, VPC details and related works are introduced. 
Section \uppercase\expandafter{\romannumeral3} details the proposed speaker anonymization system and strategy.
Datasets and experimental setup are described in Section \uppercase\expandafter{\romannumeral4}.
Section \uppercase\expandafter{\romannumeral5} presents the experimental results and analysis.
We conclude in Section \uppercase\expandafter{\romannumeral6}.


\begin{figure*}[ht]
  \centering
  \includegraphics[width=14cm]{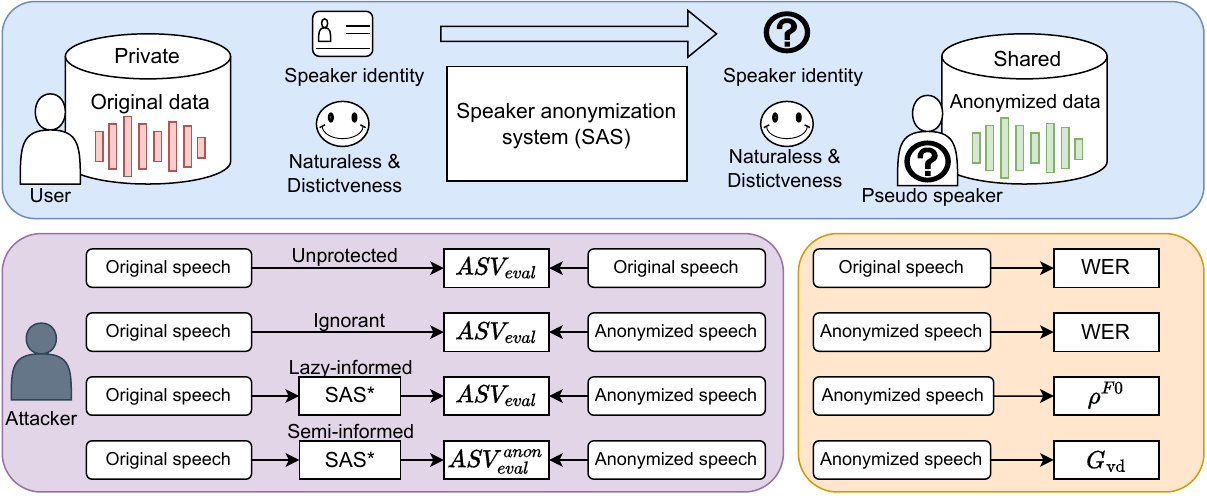}
  \caption{An overview of the VPC official design~\cite{vpc2022eval}. Users anonymize the original speech to hide their personal identity while preserving content, naturalness, and distinctiveness, as shown in the blue box. The purple box supposes attackers use different ASV systems and knowledge of the anonymization system to re-identify the original speaker's identity in different attack scenarios. $ASV_{eval}$ and $ASV_{eval}^{anon}$ represent the ASV model trained on original speech data and anonymized data, respectively. SAS$^*$ denotes a speaker anonymization system similar to SAS but lacking accurate parameters.}
  \vspace{-15pt}
  \label{fig:fig_vpc}
\end{figure*}

\section{Related works}
This section serves as a comprehensive introduction to the official design of VPC, setting the stage for this study. It includes definitions of specific scenarios, attack models, and evaluation metrics. Meanwhile, we also overview the two classes of speaker anonymization approaches mentioned and their limitations.

\vspace{-5pt}
\subsection{VoicePrivacy Challenge}\label{vpc_intro}
Speaker anonymization in VPC aims to protect users who share personal speech data, as shown in Figure~\ref{fig:fig_vpc}. The user shared data anonymized by speaker anonymization system (SAS) and expects to be against malicious re-identify from attackers, as shown in the lower of Figure~\ref{fig:fig_vpc}.
To measure such a system, VPC gives four official metrics:
\begin{itemize}
\item \textit{Privacy metric}:
The privacy protection ability of the anonymization system is assessed with the ASV model provided by VPC, in terms of the equal error rate (EER) is computed as the primary privacy metric. 
An ideal anonymization system should achieve high EERs (close to 50\%) to protect users' privacy information \cite{nn5}.
\item \textit{Intelligibility metric}:
To assess the preservation of linguistic content and intelligibility in anonymized speech, the Word Error Rate (WER) serves as a valuable metric computed by an official ASR model. 

\item \textit{Intonation metric}:
To measure how well anonymization preserves the intonation of the original speech, the pitch correlation metric $\rho^{\textrm{F0}}$ is computed~\cite{pitchcorrelation}. $\rho^{\textrm{F0}}$ is the Pearson correlation~\cite{pearson} between the pitch sequences of original speech and anonymized speech. 
\item \textit{Voice distinctiveness metric}:
The gain of voice distinctiveness metric ($G_{\textrm{vd}}$) aims to assess the voice distinctiveness preservation of anonymized speech~\cite{gvd1,gvd2}. It relies on two voice similarity matrices. A voice similarity matrix $M=(M(i,j))_{1 \leq i \leq N, 1 \leq j \leq N}$ is computed for N speakers and similarity values $M=(M(i,j))$ computed for speakers $i$ and $j$ as follows:

\begin{align}
\resizebox{0.9\linewidth}{!}{$M(i, j)=\operatorname{sigmoid}\left(\frac{1}{n_i n_j} \sum_{\substack{1 \leq k \leq n_i \\ 1 \leq l \leq n_j \\ k \neq l \text { if } i=j}} \operatorname{LLR}\left(x_k^{(i)}, x_l^{(j)}\right)\right)$},
\end{align}
where $\operatorname{LLR}(x_k^{(i)}, x_l^{(j)})$ is the log-likelihood-ratio obtained by comparing the $k$-th segment from the $i$-th speaker with the $l$-th segment from the $j$-th speaker and $n_i$ and $n_j$ are the numbers of utterances for speaker $i$ and speaker $j$, respectively. The log-likelihood ratio is obtained by the ASV model. $M_{oo}$ (for \textbf{o}riginal data) and $M_{aa}$ (for \textbf{a}nonymized data) are prepared for computing the diagonal dominance $D_{\text {diag }}(M)$, which is the absolute difference between the mean values of diagonal and off-diagonal elements, formulated as follows:
\begin{equation}
    D_{\textrm{diag }}(M)=\left|\sum_{1 \leq i \leq N} \frac{M(i, i)}{N}-\sum_{\substack{1 \leq j \leq N \\ 1 \leq k \leq N \\ j \neq k}} \frac{M(j, k)}{N(N-1)}\right|.
\end{equation}
The gain of voice distinctiveness metric $G_{\textrm{vd}}$ can be mathematically expressed as the ratio between the diagonal dominance of two matrices:
\begin{equation}
    G_{\mathrm{vd}}=10 \log _{10} \frac{D_{\mathrm{diag}}\left(M_{\mathrm{aa}}\right)}{D_{\mathrm{diag}}\left(M_{\mathrm{oo}}\right)},
\end{equation}
$G_{\mathrm{vd}}=0$ signifies that the voice distinctiveness remains unchanged after anonymization. Conversely, a $G_{\mathrm{vd}}$ gain value above or below 0 indicates an average increase or decrease in voice distinctiveness, respectively.

\item \textit{Subjective metric:} Mean Opinion Score (MOS) and Similarity Mean Opinion Score (SMOS) are employed to evaluate the naturalness and privacy protection of the anonymized samples. Participants are tasked with listening to the provided speech and evaluating its naturalness on a 5-point scale: 1 - bad, 2 - poor, 3 - fair, 4 - good, and 5 - excellent.
\end{itemize}

The attack scenarios are assumed under the two attack models shown in the lower left of Figure~\ref{fig:fig_vpc}. The attackers are able to access a few original or anonymized utterances for each speaker and have different levels of knowledge about the SAS. The details of attack scenarios are as follows:

\textit{Unprotected}: 
The protection of the user's personal speech data through speaker anonymization is not used; potential attackers can directly verify the user's identity by employing an ASV model ($ASV_{eval}$) trained on the original data.

\textit{Ignorant}:
The user's personal speech data is protected by the speaker anonymization system, but attackers are unaware and still verify the user's identity by $ASV_{eval}$.

\textit{Lazy-informed}: Attackers are known about SAS and perform a similar system, denoted as SAS*, to anonymize their enrollment data to verify the user's identity by $ASV_{eval}$. 

\textit{Semi-informed}:
Going a step further, attackers upgrade their attack model by fine-tuning $ASV_{eval}$ with anonymized speech to improve the verification ability. And then, attackers employ the upgraded model ($ASV_{eval}^{anon}$) to verify the user's identity.

\vspace{-5pt}
\subsection{Signal-processing Based Speaker Anonymization}
The majority of signal processing-based approaches do not require training data and directly manipulate instantaneous speech characteristics. The current mainstream approach primarily focuses on manipulating formant frequency and F0.
A typical approach involves using McAdams coefficients~\cite{mcadams1984spectral} to introduce randomized shifts to the formant frequencies, effectively distorting the spectral envelope to anonymize the original speaker's identity~\cite{sig1}. Building on this McAdams coefficients approach, Gupta \textit{et al.} further extends the distortion of the spectral envelope by amplifying the width of formant peaks~\cite{sig2}.
Apart from formant shifts, another approach also focuses on F0 trajectories and manipulates them based on functional data analysis techniques~\cite{sig3}.

While these methods can perceptually manipulate the speech signal, they are less effective at protecting the speaker's identity completely. Powerful attackers can potentially restore the original speech after making a reasonable number of attempts~\cite{vpc2020eval, vpc2022eval}. This limitation indicates the necessity for more robust and advanced speaker anonymization techniques to ensure better privacy protection.

\vspace{-5pt}
\subsection{Neural Voice Conversion Based Speaker Anonymization}
As another solution for VPC, neural voice conversion-based speaker anonymization is commonly used in most approaches. This method generally outperforms signal-processing-based approaches in terms of privacy and utility metrics.
A typical approach involves using a pre-trained ASV model to extract speaker identity representations~\cite{xvectoranonymization}, like x-vectors~\cite{xvector} or d-vectors~\cite{dvector}. At the same time, an ASR model is used to extract linguistic content representations. Subsequently, the original speaker representation is replaced by averaging a set of candidate speaker representations from an external pool. Their farthest representations are chosen based on cosine distance to select candidate speakers, ensuring privacy protection. Finally, the averaged speaker representation, linguistic content, and F0 are processed by the neural voice conversion model, generating an anonymized mel-spectrogram. A vocoder then generates anonymized speech from the anonymized mel-spectrogram.

In current mainstream studies, neural voice conversion for speaker anonymization follows a similar framework to the baseline approach but has seen improvements in various aspects. One significant area of improvement lies in the anonymization process, where researchers have modified the averaging anonymization method to enhance its privacy protection capabilities.
Indeed, some studies~\cite{adv_anon,gan_anon} have employed adversarial training strategies to generate a pseudo speaker vector from a pool of speaker vectors to suppress speaker-related information. On the other hand, alternative approaches~\cite{pca_anon, xvector_j,nn10} model a speaker-level latent space and utilize techniques such as Principal Component Analysis (PCA) or SVD to obtain an anonymized vector for each speaker.

\begin{figure*}[ht]
  \centering
  \includegraphics[width=13cm]{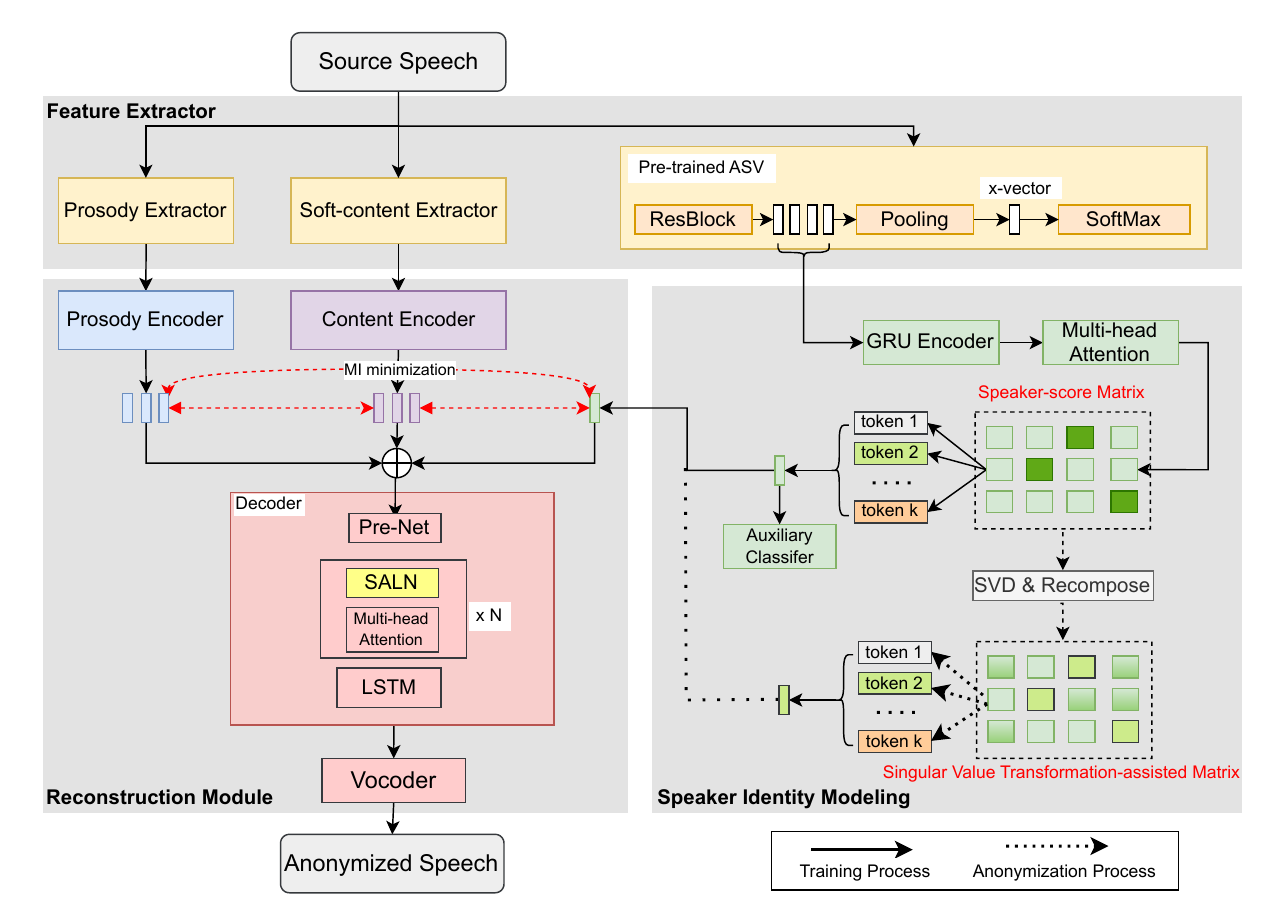}
  \caption{An overview of our proposed anonymization system. The black dashed line delineates the anonymization process, while the red dashed line represents the Mutual Information (MI) minimization of linguistic content, prosody, and speaker identity. SALN and SVD represent style-adaptive layer normalization and singular
value decomposition, respectively.} \vspace{-15pt}
  \label{fig:fig_model}
\end{figure*}
The second part focuses on reducing the residual speaker information in the linguistic content feature or F0 feature. In~\cite{asr_anon}, researchers believe that the bottleneck feature extracted from the ASR model may still contain residual speaker information and use text transcription recognized from the ASR model to replace the bottleneck feature. On the other hand, several approaches modify the F0 trajectories using transformations towards a target pitch or by adding noise to cancel the possibility of speaker identity leakage~\cite{meyer2023prosody,champion2021study,nn8}.
In addition to these approaches, Miao \textit{et al.}~\cite{nn6,miao2023speaker} used a HuBERT-based soft content encoder to extract language-independent linguistic content which can anonymize speech from any language without relying on text labels or other language-specific resources.

Each approach requires a pre-trained ASV model to extract utterance-level speaker identity representation and external speaker vector pool. However, the extracted speaker representation can not adapt to the voice conversion model well due to the ASV model and the conversion model being optimized by the different loss objectives~\cite{dgc2022}. It will lead to a large degradation in naturalness if directly averaging or modifying the extracted speaker representation.
By contrast, an alternative approach was introduced by Yao \textit{et al.}~\cite{nn7}, wherein a look-up-table (LUT) based anonymization technique was employed to generate speaker embeddings. Ranked first in VPC 2022, this method obviates the necessity for a pre-trained ASV model and the external speaker vector pool typically required by other approaches. 
Nevertheless, one notable limitation of this approach is the degradation of the speaker's distinctiveness. Moreover, empirical evaluations in VPC have revealed that the approach still exhibits certain shortcomings in terms of naturalness~\cite{vpc2022eval}.

\section{Methodology}\label{sec:method}
In this section, we introduce how the proposed system anonymizes the original speech to mitigate the problems with existing approaches. We will start with a system overview, followed by an introduction to speaker identity representation modeling. Next, we will describe our proposed anonymization strategy, and finally, we will define our training objectives.

\vspace{-5pt}
\subsection{System Overview}
Fig~\ref{fig:fig_model} illustrates the framework of our proposed anonymization system. Our framework consists of three modules: feature extractor, speaker identity modeling module, and reconstruction module.
Before the training stage, three features are extracted: prosody representation, soft content representation, and a sequence of frame-level speaker vectors. The feature extractor consists of two pre-trained models: a HuBERT-based soft content extractor~\cite{soft-hubert}, an ECAPA-TDNN speaker extractor~\cite{ecapatdnn}, and one prosody extractor. The pre-trained model is frozen during training.
The soft content extractor~\cite{soft-hubert} is obtained through the process of fine-tuning a pre-trained HuBERT-Base model~\cite{hsu2021hubert}, with a specific focus on obtaining fine-grained continuous context representations derived from discrete units. The linguistic content feature extracted from this model can effectively mitigate ambiguous mispronunciations and increase the overall naturalness of anonymized speech~\cite{soft-hubert}. On the other hand, the ECAPA-TDNN is responsible for extracting a frame-level representation related to the speaker's characteristics. Regarding the prosody extraction, the YAAPT algorithm~\cite{yaapt} is employed to extract energy and F0 from the input speech signal as prosody feature.

To model the speaker identity representation, we utilize a speaker identity modeling module to extract a global time-variant representation. This module transforms a sequence of frame-level speaker vectors extracted from the pre-trained ECAPA-TDNN into an utterance-level speaker-related representation $S$ through the attention mechanism.
For linguistic content and prosody modeling, the extracted soft content and prosody features serve as the input of the soft content encoder and prosody encoder to obtain hidden linguistic content representation $C=\{c_1,c_2,...,c_{T/2}\}$ and prosody representation $P=\{p_1,p_2,...,p_{T}\}$, where $T$ is frame length. 
Subsequently, the decoder uses the hidden linguistic, prosody, and speaker representation to predict the mel-spectrogram. Finally, a pre-trained vocoder reconstructs the speech waveform from the predicted mel-spectrogram.

\begin{figure}[ht]
  \centering
  \includegraphics[width=6cm]{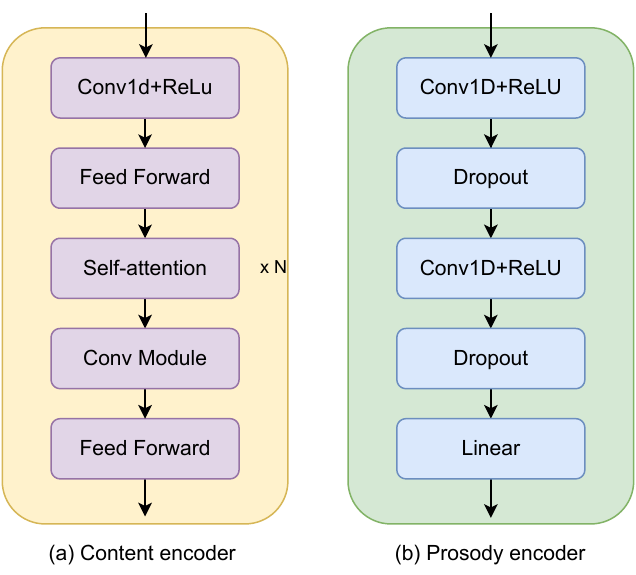}
  \caption{The overall architecture for the content encoder and prosody encoder.} \vspace{-10pt}

  \label{fig:fig_encoder}
\end{figure}

The details of the content encoder and the prosody encoder are illustrated in Fig~\ref{fig:fig_encoder}. The content encoder consists of a 1D convolutional network and four conformer blocks.
The prosody encoder consists of a 2-layer 1D convolutional network with ReLU activation, each followed by the dropout layer, and an extra linear layer to project the extracted F0 and energy into the hidden representations.
Both encoders encode the soft content feature and prosody to the same dimension as the speaker identity representation, aiming for mutual information minimization and addition between different representations.

Details of speaker identity representation modeling and anonymization process will be expounded in subsequent subsections. The final objective function to train our anonymization system will also be introduced in Section~\ref{loss}.

\vspace{-5pt}
\subsection{Speaker Identity Representation Modeling}

The role of speaker identity representation in speaker anonymization assumes paramount significance. Its importance can be categorized into two aspects: firstly, the pursuit of attaining an optimal representation of the speaker's identity, and secondly, the concealment of the speaker's identity. An ideal anonymized speaker representation should: 1) cancel the original speaker identity information, 2) keep a unique speaker identity to maintain the diversity of anonymized speakers, and 3) exclude any residual speaker irrelevant information.
Traditionally, the prevailing approach assumes speaker identity as a global representation and directly extracts an utterance-level speaker vector, commonly known as the x-vector~\cite{xvector}. Unfortunately, these approaches neglect other valuable time-variant information within speaker identity. Speakers indeed can modulate their timbre using diverse speaking techniques, skillfully utilized to convey precise semantic nuances~\cite{jiang2023mega, mcadams2013musical}. 

To avoid the loss of speaker identity information and reduce the residual speaker irrelevant information, we model the speaker identity representation in three steps:

(i). We extract the frame-level speaker vector from the pre-trained ASV model instead of the conventional utterance-level vector. As shown in Fig.~\ref{fig:fig_model}, the pre-trained ASV model consists of several ResNet blocks~\cite{ecapatdnn}, a pooling layer, and a fully connected layer. We extract the output of the last ResNet block as speaker-related vectors, which retain time dimension. 

(ii). The frame-level speaker vector serves as the input to the GRU encoder, designed to capture both time-invariant and other time-variant speaker identity information, we utilize the output from the GRU layer $\widetilde{S}=\{\widetilde{s}_1, \widetilde{s}_2,...,\widetilde{s}_T\}$ as the attention query. The GRU encoder consists of three convolutional layers and a GRU layer.

(iii). A multi-head attention layer is employed to capture the correlation between $\widetilde{S}$ and speaker-related tokens $K$ while generating the speaker-score matrix $\mathbf{O}$. 
It can be denoted by:
\begin{align}
    \mathbf{O} &=\textrm{softmax}(\frac{\widetilde{S} K^\textrm{T}}{\sqrt{d_k}}), 
\end{align}
where $\mathbf{O} \in \mathbb{R}^{i*j}$, $i$ and $j$ represent the $i$th attention head and the $j$th token and $d_k$ represents the dimensional of the attention query and key. $K$ are randomly initialized and can be made learnable.
This configuration empowers the attention layer to capture salient features and align the most relevant information across the speaker identity representation. Once we obtain the speaker-score matrix, we can generate a fixed-dimension speaker vector in a multi-head manner:
\begin{equation}
    S=\sum_{i=1}^{n} \mathbf{O}_i K_i, 
\end{equation}
where $n$ represents the total head quantity of the attention layer. We eventually get the final speaker identity representation by multiplying the speaker-score matrix and speaker-related tokens. 

Conventionally, the speaker identity representation is provided to the decoder simply through either the concatenation or the summation of the linguistic content and prosody information. In contrast, in the decoder phase, we adopt the novel paradigm of Style-Adaptive Layer Normalization (SALN)~\cite{min2021metastylespeech} to enhance the control of the speaker identity. Suppose $H = {h_1,h_2,...,h_D}$ is the input hidden feature, where $D$ is the dimension of the input hidden feature. Speaker identity representation is used to predict the gain and bias:
\begin{equation}
    \textrm{SALN}(H,S)=g(S) \cdot \textrm{Norm}(H)+b(S), 
\end{equation}
$g(\cdot)$ and $b(\cdot)$ represent the affine layer to obtain the gain and bias of $H$. The SALN can adaptively perform scaling and shifting of the normalized input features based on the speaker identity representation.

\vspace{-10pt}
\begin{figure}[ht]
  \centering
  \includegraphics[width=8cm]{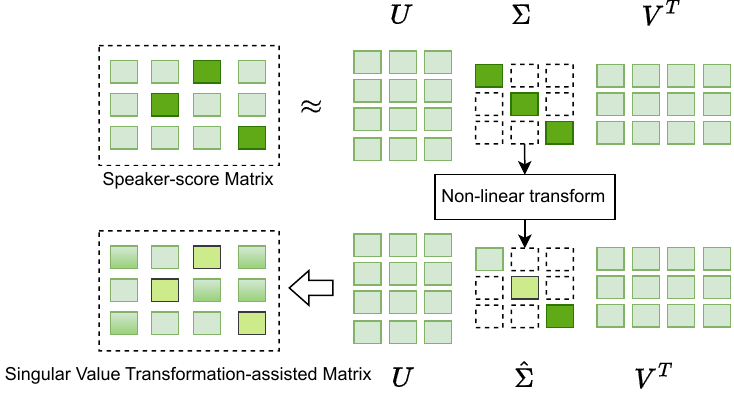}
  \caption{The generation process of the singular value transformation-assisted matrix using SVD. The upper portion illustrates the decomposition process of the original speaker-score matrix, while the lower portion is the re-composition process of the singular value transformation-assisted matrix.}
\vspace{-10pt}
  \label{fig:svd}
\end{figure}

\vspace{-15pt}
\subsection{Anonymization Strategy}

In the preceding section, we introduce the speaker identity representation generated by the multiplication between the speaker-score matrix and the speaker-related tokens. The speaker-related tokens are similar to a soft speaker label, and different speaker-score matrices distinctly correspond to different speakers. Therefore, the singular value transformation-assisted matrix is generated via matrix transformation using SVD. Fig.~\ref{fig:svd} shows the generation process of the singular value transformation-assisted matrix. It can be divided into a three steps process:

(i). The speaker-score matrix $\mathbf{O}$ is decomposed by SVD into two orthonormal eigenvectors matrices, i.e., $\mathbf{U}$ and $\mathbf{V}$, and a diagonal singular values matrix $\mathbf{\Sigma}$. $\mathbf{U}$ and $\mathbf{V}$ are the orthonormal eigenvectors of $\mathbf{O}\mathbf{O}^\textrm{T}$ and $\mathbf{O}^\textrm{T}\mathbf{O}$, respectively, and $\mathbf{\Sigma}$ consists of the square roots of the eigenvalues of $\mathbf{O}\mathbf{O}^\textrm{T}$.

(ii). The diagonal of singular matrix $\mathbf{\Sigma}$ is composed of $i$ singular values, representing the importance of the original speaker-score matrix in each dimension. We use a logarithmic transformation to decrease the magnitude of the larger singular value while simultaneously enhancing the smaller singular values. This transformation effectively reduces the gap between each singular value, thereby adjusting the significance of the singular value in each dimension. As a result, the value of the speaker score matrix is significantly changed.

(iii). After the logarithmic transformation, we recompose the speaker-score matrix using $\mathbf{U}$, $\mathbf{V}$, and the transformed $\hat{\mathbf{\Sigma}}$ to generate the singular value transformation-assisted matrix.

\begin{algorithm}
\caption{Anonymization process through SVD}
\begin{algorithmic}[1]
\REQUIRE $\mathbf{O}$: speaker-score matrix, $\mathbf{O} \in \mathbb{R}^{i*j}$ 
\REQUIRE $\alpha$: similarity threshold 
\WHILE{$\textrm{cos}(\mathbf{O}, \mathbf{O}_a) > \alpha$}
    \STATE Initialize $\mathbf{O}_a \leftarrow \mathbf{O}$
    \STATE Decompose speaker-score matrix by $\mathbf{O} \rightarrow U\Sigma V^\textrm{T}$
    \STATE Update singular value matrix of $\mathbf{O}$ by non-linear transform: $\Sigma \rightarrow \hat{\Sigma}$
    \STATE Recompose a new speaker-score matrix: $U\hat{\Sigma}V^\textrm{T} \rightarrow \hat{\mathbf{O}}$
    \STATE Find the most different score index for each attention head $e_i = \textrm{argmax} (\hat{\mathbf{O}}(i,:) - \mathbf{O}(i,:))$
    \STATE Get singular value transformation-assisted matrix $\mathbf{O}_a$ by updating $\mathbf{O}$: $\mathbf{O}_a(i, j) \leftarrow \mathbf{O}(i,e_i)$
\ENDWHILE
\RETURN Singular Value Transformation-assisted Matrix $\mathbf{O}_a$
\end{algorithmic}
\end{algorithm}
Based on SVD, the singular value transformation-assisted matrix is compared with the original speaker-score matrix to select the score index corresponding to the most different scores for each attention head. Then, we update the original speaker-score matrix relying on the selected scores and repeat the above process until the cosine similarity is less than the threshold $\alpha$. Algorithm 1 outlines the anonymization process integrated with SVD.

\vspace{-10pt}
\subsection{Five Specifically-Designed Objectives}\label{loss}
During the model training, we define a reconstruction loss $\mathcal{L}_\textrm{rec}$ that constrains the mel-spectrogram $X$ to reconstruct itself based on linguistic content $C$, prosody representation $P$, and speaker identity representation $S$. The reconstruction loss can be expressed as follow:
\begin{equation}
    \mathcal{L}_\textrm{rec}(\theta_{E_c},\theta_{E_p},\theta_{E_s}, \theta_{\textrm{Dec}})=||\textrm{Dec}(C,P,S)-X||, 
\end{equation}
where $\theta$ denotes the parameter of the corresponding module. Thus, we can reconstruct the original mel-spectrogram by giving the linguistic, prosody, and speaker identity representation.

On the other hand, we introduce MI as the correlation metric to achieve more proper disentanglement of linguistic content, prosody, and speaker identity. Given the variables $u$ and $v$, MI is Kullback-Leibler (KL)~\cite{kl} divergence $D_{\textrm{KL}} (\mathbf{P}(u,v); \mathbf{P}(u)\mathbf{P}(v))$ between variables' joint and marginal distributions. We adopt vCLUB~\cite{vclub} to compute the upper bound of MI as:

\begin{align}
    \mathcal{L}_\textrm{mi}(C,S)=\mathbb{E}_{\mathbf{P}(C,S)}&[log Q_{\theta_{C, S}}\left(c_{t} \mid s_{t} \right)] \notag\\ &- 
    \mathbb{E}_{\mathbf{P}(C)\mathbf{P}(S)}[log Q_{\theta_{C, S}}\left(c_{t} \mid s_{t} \right)], 
\end{align}
\begin{align}
    \mathcal{L}_\textrm{mi}(P,S)=\mathbb{E}_{\mathbf{P}(P,S)}&[log Q_{\theta_{P, S}}\left(p_{t} \mid s_{t} \right)] \notag\\ &- 
    \mathbb{E}_{\mathbf{P}(P)\mathbf{P}(S)}[log Q_{\theta_{P, S}}\left(p_{t} \mid s_{t} \right)], 
\end{align}

\begin{align}
    \mathcal{L}_\textrm{mi}(C,P)=\mathbb{E}_{\mathbf{P}(C,P)}&[log Q_{\theta_{C, P}}\left(c_{t} \mid p_{t} \right)] \notag\\ &- 
    \mathbb{E}_{\mathbf{P}(C)\mathbf{P}(P)}[log Q_{\theta_{C, P}}\left(c_{t} \mid p_{t} \right)], 
\end{align}
where $Q_{\theta_{(\cdot), (\cdot)}}$ is the variational approximation of ground-truth posterior and can be parameterized by the neural network $Q_{\theta}$.

Furthermore, we add an auxiliary classification constraint to ensure the speaker-score matrix models the speaker-related information. The classifier takes the final speaker identity representation as input and predicts its speaker ID. The auxiliary classification constraint is defined by
\begin{equation}
    \mathcal{L}_{\textrm{cls}}=\mathbb{E}[-log(C_\theta(I\mid S))], 
\end{equation}
where $C_\theta$ and $I$ represent auxiliary classifier and speaker ID, respectively. Without the auxiliary classifier, the speaker-score matrix could potentially encompass irrelevant information about the speaker's identity.
Finally, the overall objective $\mathcal{L}$ is the minimization of five weighted loss function:
\begin{align}\label{total_loss}
    \mathcal{L} = \mathcal{L}_\textrm{rec} + \lambda\mathcal{L}_\textrm{mi}(C,S)+\lambda\mathcal{L}_\textrm{mi}(P,S)+\lambda\mathcal{L}_\textrm{mi}(C,P)+\mathcal{L}_{\textrm{cls}}, 
\end{align}
where $\lambda$ is a constant weight to control how MI loss enhances the disentanglement and we set $\lambda=0.01$. 

\begin{table*}[ht]
\caption{Privacy and utility evaluation results of our proposed anonymization system. EER achieved by $ASV_{\textrm{eval}}$ and $ASV_{\textrm{eval}}^{\textrm{anon}}$ on different attack scenarios are used as the privacy metrics. Other metrics are utilized to measure the utility.}\label{tab:details}
\renewcommand\arraystretch{1.1}
\resizebox{1.0\linewidth}{!}{
\begin{tabular}{cccccccc}
\hline
\multirow{2}{*}{Dataset}              & \multirow{2}{*}{Gender} & \multirow{2}{*}{Weight} & \multirow{2}{*}{\makecell[c]{Ignorant\\EER $\uparrow$} } & \multirow{2}{*}{\makecell[c]{Lazy-informed\\EER $\uparrow$} } & \multicolumn{1}{l}{\multirow{2}{*}{\makecell[c]{Semi-informed\\EER $\uparrow$} }} & \multicolumn{1}{l}{\multirow{2}{*}{$\rho^{\textrm{F0}}$ $\uparrow$}} & \multirow{2}{*}{$G_{\textrm{vd}}$ $\uparrow$ } \\
                                      &                         &                         &                       &                       & \multicolumn{1}{l}{}                      & \multicolumn{1}{l}{}                   &                      \\ \hline
\multirow{2}{*}{LibriSpeech-dev}      & female                  & 0.25                    & 59.14                  & 53.39                  & 41.20                                      & 0.83                                   & -0.75                 \\ \cline{2-3}
                                      & male                    & 0.25                    & 58.79                  & 50.28                  & 44.53                                      & 0.84                                   & -0.89                 \\ \hline
\multirow{2}{*}{VCTK-dev (different)} & female                  & 0.20                    & 58.70                  & 43.45                  & 38.17                                      & 0.82                                   & -1.26                 \\ \cline{2-3}
                                      & male                    & 0.20                    & 58.75                  & 41.39                  & 37.82                                      & 0.85                                   & -1.65                 \\ \hline
\multirow{2}{*}{VCTK-dev (common)}    & female                  & 0.05                    & 55.18                  & 47.32                  & 36.49                                      & 0.81                                   & -1.42                 \\ \cline{2-3}
                                      & male                    & 0.05                    & 56.37                  & 43.65                  & 32.94                                      & 0.82                                   & -1.23                 \\ \hline
\multicolumn{3}{c}{Weighted average dev}                                                      & \textbf{57.23}                  & \textbf{47.44}                  & \textbf{40.11}                                      & \textbf{0.83}                                   & \textbf{-1.12}                 \\ \hline

\multirow{2}{*}{LibriSpeech-test}      & female                  & 0.25                    & 56.64                  & 48.45                  & 41.79                                      & 0.84                                   & -0.56                 \\ \cline{2-3}
                                      & male                    & 0.25                    & 52.75                  & 49.15                  & 42.33                                      & 0.85                                   & -0.84                 \\ \hline
\multirow{2}{*}{VCTK-test (different)} & female                  & 0.20                    & 50.45                  & 41.57                  & 39.73                                      & 0.83                                   & -1.41                 \\ \cline{2-3}
                                      & male                    & 0.20                    & 48.35                  & 42.89                  & 36.26                                      & 0.84                                   & -1.15                 \\ \hline
\multirow{2}{*}{VCTK-test (common)}    & female                  & 0.05                    & 52.28                  & 40.92                  & 36.15                                      & 0.83                                   & -1.27                 \\ \cline{2-3}
                                      & male                    & 0.05                    & 51.07                  & 41.45                  & 33.98                                      & 0.82                                   & -1.03                 \\ \hline
\multicolumn{3}{c}{Weighted average test}                                                      & \textbf{51.28}                  & \textbf{45.41}                  & \textbf{39.74}                                      & \textbf{0.84}                                   & \textbf{-0.98}                 \\ \hline
\end{tabular}
}\vspace{-15pt}
\end{table*}

\vspace{-10pt}
\section{Experimental Setup}
To evaluate the performance of our proposed anonymization system under all the attack scenarios, i.e., Ignorant, Lazy-informed, and Semi-informed, we follow the VPC evaluation plan~\cite{vpc2022eval} described in Section~\ref{vpc_intro} and conduct extensive experiments based on the official datasets. In this section, the datasets for model training and evaluation will be introduced. Besides, implementation details and additional subjective evaluation metrics will also be presented.
\vspace{-15pt}
\subsection{Datasets}
The datasets used for training developments and testing are shown in Table~\ref{tab:datasets}. Specifically, the soft-content extractor\footnote{\url{https://github.com/bshall/hubert}} is trained on \textit{LibriSpeech-train-clean-100} and \textit{LibriSpeech-train-other-500} datasets~\cite{librispeech}.
The ECAPA-TDNN speaker verification model is pre-trained on the \textit{VoxCeleb2} dataset~\cite{vocceleb}, employing the SpeechBrain toolkits\footnote{\url{https://github.com/speechbrain/speechbrain}} to extract frame-level speaker vectors.
Both the acoustic model and the vocoder are trained on \textit{LibriTTS-train-clean-100} and \textit{LibriTTS-train-other-500}~\cite{libritts} datasets. After training, our anonymization system is evaluated on the official VPC development and test datasets~\cite{vpc2022eval}, which contain several female and male speakers from \textit{LibriSpeech dev-clean}, \textit{LibriSpeech test-clean}, and \textit{VCTK} corpora~\cite{vctk}. Both are split into trial and enrollment subsets. For the Ignorant and Lazy-informed scenarios, we use $ASV_\textrm{eval}$ system provided by the VPC for evaluation. In the Semi-informed scenario, we follow the VPC official process to retrain the $ASV_\textrm{eval}^{\textrm{anon}}$ system using our anonymized speech data. 

\begin{table}[h]
\centering
\caption{Datasets used for training, development and testing}\label{tab:datasets}
\begin{tabular}{l|c|c|c}
\hline
                             & Dataset                     & Speaker & Utterances \\ \hline
\multirow{5}{*}{Training}    & VoxCeleb-2                  & 6,112   & 1,281,762  \\
                             & LibriSpeech train-clean-100 & 251     & 28,539     \\
                             & LibriSpeech train-other-500 & 1,166   & 148,688    \\
                             & LibriTTS train-clean-100    & 247     & 33,236     \\
                             & LibriTTS train-other-500    & 1,160   & 205,044    \\ \hline
\multirow{2}{*}{Development} & LibriSpeech dev-clean       & 69      & 2,321      \\
                             & VCTK-dev                    & 30      & 11,972     \\ \hline
\multirow{2}{*}{Testing}        & LibriSpeech test-clean      & 69      & 1,943      \\
                             & VCTK-test                   & 30      & 12,048     \\ \hline
\end{tabular}
\end{table}

\vspace{-15pt}
\subsection{Setup}
To clearly demonstrate the effectiveness of our proposed anonymization system, we compare it with various baseline systems. The first two baseline systems are denoted as \textbf{B1.a} and \textbf{B1.b}, which are the VPC official baseline systems~\cite{vpc2022eval}. 
Additionally, we compare the proposed anonymization system with the first-ranking anonymization system in the VPC 2022, referred to as \textbf{ASV-Free}, which uses LUT to replace the x-vector pool and combines two types of speaker embedding to achieve speaker anonymization~\cite{nn7}. 
We also consider an upgraded version of \textbf{ASV-Free} called \textbf{Formant}~\cite{yao2023distinguishable}, which models speaker identity based on formant and F0 rather than LUT, serving as another baseline system.

In our proposed anonymization system, the speaker encoder takes a sequence of 192-dimensional speaker vectors with a time dimension. We configure the head of the multi-head attention in the speaker encoder to be 8, and the token number is set to 1000. The content encoder, prosody encoder, speaker encoder, and decoder are optimized by Equation (~\ref{total_loss}). 
During the training stage, the speech data's sample rate is 16kHz, and it is transformed into 80-dimensional mel-spectrograms using a frame size of 50ms and a frame-shift of 10ms. We utilize a batch size of 32 and the Adam optimizer with a learning rate 2e-4. The model is trained for 500k steps.
The structure and training strategy of the vocoder is the same as the official implementation\footnote{\url{https://github.com/jik876/hifi-gan}} of HiFi-GAN~\cite{hifigan}. Regarding the anonymization process, we set the similarity threshold $\alpha=0.75$.

\begin{table*}[ht]
\caption{Averaged results over baselines and proposed approaches on VPC development and test datasets. The best scores in the column of the test datasets are given in bold. For EER, closer to 50\% is better. For WER, lower is better. For $\rho^{\textrm{F0}}$ and $G_{\textrm{vd}}$, higher is better. }\label{tab:compare}
\renewcommand\arraystretch{1.2}
\resizebox{1.0\linewidth}{!}{
\begin{tabular}{lcccccccccc}
\hline
\multicolumn{1}{c}{} & \multicolumn{2}{c}{\textbf{B1.a}~\cite{vpc2022eval}} & \multicolumn{2}{c}{\textbf{B1.b}~\cite{vpc2022eval}} & \multicolumn{2}{c}{\textbf{ASV-Free}~\cite{nn7}} & \multicolumn{2}{c}{\textbf{Formant}~\cite{yao2023distinguishable}} & \multicolumn{2}{c}{\textbf{Proposed}} \\ \cline{2-11} 
\multicolumn{1}{c}{} & dev             & test            & dev             & test            & dev               & test              & dev               & test             & dev               & test              \\ \hline
Ignorant EER by $ASV_{\textrm{eval}}$ $\uparrow$         & 53.14           & 50.29           & 53.91           & \textbf{52.14}           & 52.11             & 51.27             & 49.55             & 48.63            & 57.23             & 51.28             \\
Lazy-informed EER by $ASV_{\textrm{eval}}$ $\uparrow$    & 32.12           & 32.82           & 27.39           & 27.51           & 40.83             & 39.77             & 41.22             & 40.76            & 47.44             & \textbf{45.41}             \\
Semi-informed EER by $ASV_{\textrm{eval}}^{\textrm{anon}}$ $\uparrow$    & 11.74           & 11.81           & 9.93            & 9.18            & 31.73             & 30.15             & 36.15             & 35.37            & 40.11             & \textbf{39.74}             \\ \hline
WER by $ASR_{\textrm{eval}}^{\textrm{anon}}$ $\downarrow$               & 7.94            & 8.29            & 7.59            & 7.56            & 5.63              & \textbf{5.86}              & 6.95              & 6.82             & 6.51              & 6.43              \\ \hline
$\rho^{\textrm{F0}}$ $\uparrow$                    & 0.77            & 0.77            & 0.80            & 0.80            & 0.72              & 0.70              & 0.80              & 0.81             & 0.83              & \textbf{0.84}              \\ \hline
$G_{\textrm{vd}}$ $\uparrow$               & -9.71           & -10.15          & -6.44           & -6.44           & -18.86            & -18.69            & -4.55             & -4.92            & -1.12             & \textbf{-0.98}             \\ \hline
\end{tabular}}\vspace{-10pt}
\end{table*}

\vspace{-15pt}
\subsection{Evaluation Metrics}
In this study, we adopt the same evaluation metrics as the VPC (described in Section~\ref{vpc_intro}). The results of the objective evaluation are computed by averaging the three results obtained on LibriSpeech-*-clean, VCTK-*(common), and VCTK-* (different) with weights of 0.5, 0.1, and 0.4, respectively.
Meanwhile, we anonymize 40 randomly selected samples from the LibriSpeech-test and 40 randomly selected samples from VCTK-test datasets, encompassing four male and four female speakers, to evaluate the naturalness and privacy protection. 

\section{Experimental Results}
\subsection{Anonymization Results on Different Attack Scenarios}
Following the VPC evaluation plan configuration, we first investigate the privacy protection performance of our proposed anonymization system in various attack scenarios. Subsequently, we evaluate our anonymization system's overall intonation and distinctiveness performance. Finally, we compare the utility and intelligibility performance on different evaluation datasets.

The results of privacy protection performance in different evaluation datasets and for different genders are presented in Table~\ref{tab:details}. For the Ignorant scenario, the average EER results in both the development and test datasets are higher than 50\%, indicating that our proposed anonymization system can effectively protect personal privacy in this scenario. 
For the other two more challenging attack scenarios, where the attacker possesses a higher knowledge level of the anonymization system than the Ignorant scenario, the EER results of our proposed system show only a slight gap. Despite this, our anonymization system still achieves reasonable privacy protection performance. These results demonstrate that our proposed anonymization system minimizes speaker privacy leakages even under more challenging attacks. 

Regarding intonation and distinctiveness performance, the $\rho^{\textrm{F0}}$ metric indicates that the pitch correlation achieved 0.83 and 0.84 in the development and test datasets, respectively. These scores are significantly higher than the VPC's requested minimum average pitch correlation of 0.3.
Additionally, the distinctiveness results of our proposed system are close to 0, which indicates the original distinctiveness is effectively preserved in anonymized speakers. The results of $\rho^{\textrm{F0}}$ and $G_{\textrm{vd}}$ metrics collectively demonstrate the superiority of our proposed system in preserving both intonation and distinctiveness.

\begin{table}[ht]
\caption{WER results of our system in different datasets, which evaluated by $ASR_{\textrm{eval}}^{\textrm{anon}}$.}\label{tab:wer}
\renewcommand\arraystretch{0.5}
\resizebox{1.0\linewidth}{!}{
\begin{tabular}{lccc}
\hline
\multicolumn{2}{c}{\fontsize{4}{6}\selectfont Dataset}         & \fontsize{4}{6}\selectfont WER  & \fontsize{4}{6}\selectfont Average               \\ \hline
\multirow{2}{*}{\fontsize{4}{6}\selectfont LibriSpeech} & \fontsize{4}{6}\selectfont dev  & \fontsize{4}{6}\selectfont 3.38 & \multirow{2}{*}{\fontsize{4}{6}\selectfont 3.35} \\
                             & \fontsize{4}{6}\selectfont test & \fontsize{4}{6}\selectfont 3.32 &                       \\ \hline
\multirow{2}{*}{\fontsize{4}{6}\selectfont VCTK}        & \fontsize{4}{6}\selectfont dev  & \fontsize{4}{6}\selectfont 9.64 & \multirow{2}{*}{\fontsize{4}{6}\selectfont 9.59} \\
                             & \fontsize{4}{6}\selectfont test & \fontsize{4}{6}\selectfont 9.54 &                       \\ \hline
\end{tabular}
}
\vspace{-10pt}
\end{table}

Table~\ref{tab:wer} shows the intelligibility performance of our proposed system. We observe a gap in the WER results between the \textit{LibriSpeech} dataset and the \textit{VCTK} dataset, despite the WER results in the \textit{VCTK} dataset are already lower. This difference may be attributed to the fact that the \textit{LibriSpeech} dataset belongs to the same domain as the training dataset of the soft content extractor. The average WER results are 3.35\% and 9.59\% for both LibriSpeech and VCTK datasets, respectively. These results demonstrate that anonymized speech can effectively preserve linguistic information and maintain better intelligibility.

\begin{figure*}[ht]
  \centering
  \includegraphics[width=13cm]{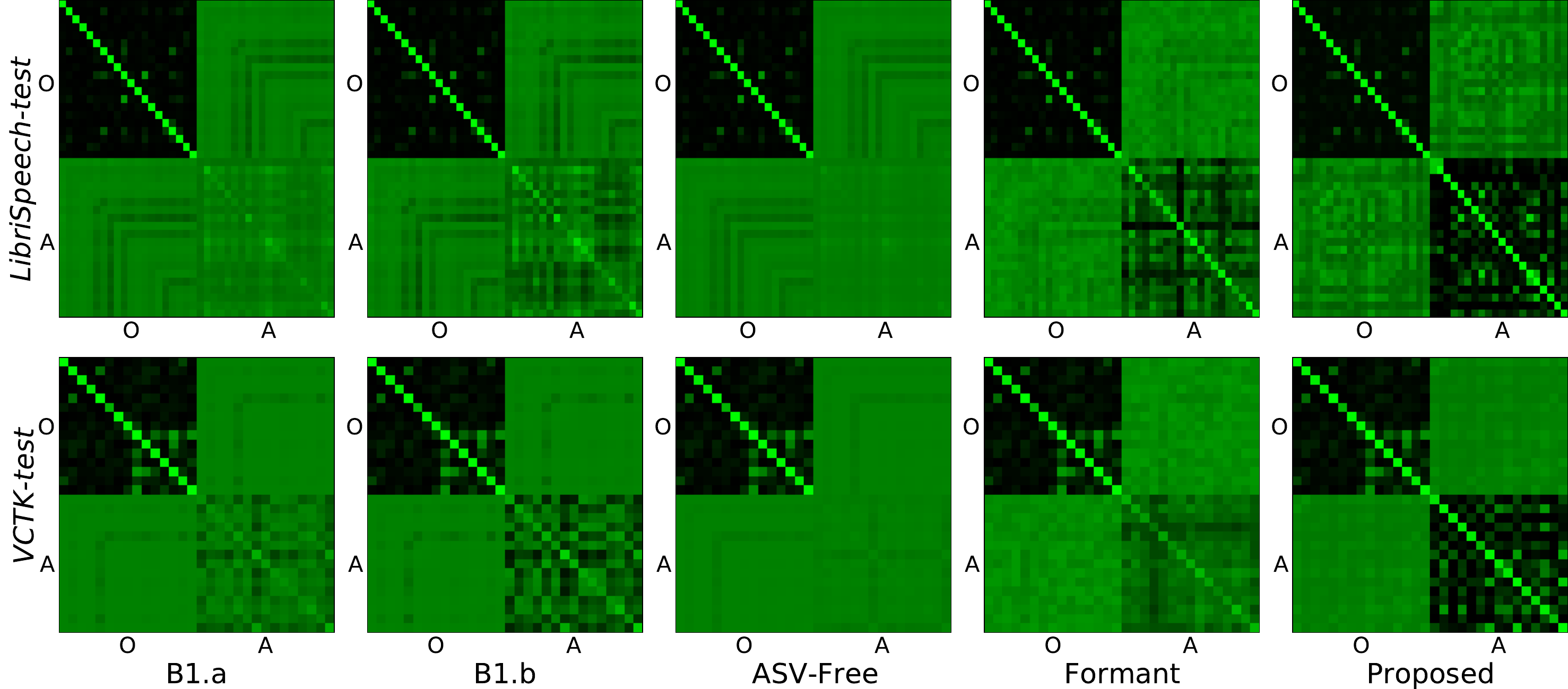}
  \caption{The visualization of the similarity matrix between the original speaker (O) and the anonymized speaker (A). In each matrix, the upper-left submatrix is $M_{oo}$, which is the similarity matrix between the original speaker and the original speaker; the upper-right $M_{oa}$ or lower-left $M_{ao}$ submatrix is the similarity matrix between the original speaker and the anonymized speaker; the lower-right submatrix $M_{aa}$ is the similarity matrix between the anonymized speaker and the anonymized speaker.}\vspace{-15pt}
  \label{fig:gvd_matrix}
\end{figure*}

\vspace{-10pt}
\subsection{Compare with Other Systems}

To further show the effectiveness of our proposed system, we compare our system with baseline systems in both subjective and objective metrics. The objective results are shown in Table~\ref{tab:compare}. 
In the Semi-informed scenario, the privacy protection performance of the B1.a and B1.b systems experiences a significant degradation, with a more substantial gap than the Ignorant or Lazy-informed scenarios. This suggests that the VPC baseline system is ineffective in preserving personal privacy when an attacker possesses knowledge of the anonymization system. On the contrary, our proposed anonymization system achieves the highest EER results among the four anonymization systems in the Semi-informed scenario, demonstrating its superior capability in protecting speaker privacy even when facing an attacker with advanced knowledge. 
Regarding the WER results, our proposed system is significantly lower than B1.a, B1.b, and Formant and is only slightly higher than ASV-Free. However, when considering the $G_{\textrm{vd}}$ results, it shows that the ASV-Free system is undesirable, as it produces a similar voice regardless of the original speaker.
Furthermore, the pitch and distinctiveness results clearly demonstrate that speech anonymized using our proposed system has a higher pitch correlation and greater diversity than the baseline systems. This indicates that our proposed anonymization system excels in preserving the speakers' original intonation and distinctiveness.

In addition to the objective distinctive metrics, we employ visualizations of the similarity matrix used in calculating the $G_{\textrm{vd}}$ metrics for the anonymized speech, as shown in Figure~\ref{fig:gvd_matrix}. The similarity matrix is computed by VPC official toolkits on LibriSpeech and VCTK test datasets. 
The distinct dominant diagonal in $M_{aa}$ reflects higher voice distinctiveness and the dominant diagonal disappears when the speaker is different. For B1.a, B1.b, and Formant, there is an unclear dominant diagonal in $M_{aa}$, indicating a deterioration in voice distinctiveness during the anonymization process. For ASV-Free, the dominant diagonal in $M_{aa}$ disappears, making the speaker indistinguishable. In contrast, the matrices for our proposed anonymization system exhibit distinct diagonals in $M_{aa}$, indicating that voice distinctiveness is successfully preserved after anonymization.

\begin{table}[h]
\centering
\caption{SMOS results of different anonymization systems. A lower score represents better privacy protection.} \label{tab:smos}
\renewcommand\arraystretch{1.5}
\resizebox{1.0\linewidth}{!}{
\begin{tabular}{lcccccc}
\hline
     & \fontsize{12}{14}\selectfont Orig & \fontsize{12}{14}\selectfont B1.a & \fontsize{12}{14}\selectfont B2.b & \fontsize{12}{14}\selectfont ASV-Free & \fontsize{12}{14}\selectfont Formant & \fontsize{12}{14}\selectfont Proposed \\ \hline
 \fontsize{12}{14}\selectfont SMOS & \fontsize{12}{14}\selectfont 0.83 & \fontsize{12}{14}\selectfont 0.52 & \fontsize{12}{14}\selectfont 0.51 &\fontsize{12}{14}\selectfont 0.39     & \fontsize{12}{14}\selectfont 0.35    & \fontsize{12}{14}\selectfont 0.32     \\ \hline
\end{tabular}}
\end{table}

\begin{figure}[h]
  \centering
  \includegraphics[width=8cm]{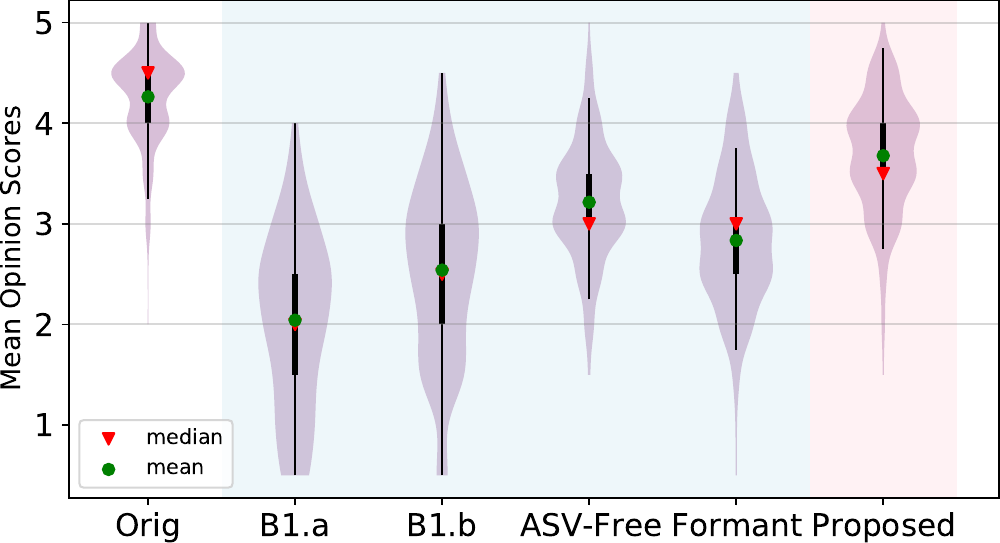}
  \caption{MOS violin-plots of the anonymized speech generated from the different anonymization systems.}
  \vspace{-10pt}
  \label{fig:mos_piano}
\end{figure}

To further analyze the naturalness of our proposed system, we conduct a subjective test and compare the MOS for naturalness between our proposed system and the baseline systems. This test allows us to gather insights into the perceived naturalness of the anonymized speech, aiding in evaluating and comparing the different systems.
Figure~\ref{fig:mos_piano} shows violin box plots of the MOS results. 
The subjective score for the VPC baseline systems is similar, with B1.b receiving higher naturalness scores than B1.a. Additionally, both ASV-Free and Formant exhibit higher naturalness scores than B1.b. However, the naturalness of anonymized speech is further improved in our proposed anonymization system, surpassing the naturalness of all the baseline systems. This implies that our proposed anonymization system is perceived to have better naturalness in the anonymized speech than other baseline systems in the subjective evaluation. Furthermore, the SMOS results are shown in Table~\ref{tab:smos}. Our proposed approach achieves an SMOS of 0.32, representing the lowest result compared to other baseline systems. This further demonstrates the superior privacy protection of our proposed approach compared to the baseline system, aligning with the same conclusion as the EER results.

\vspace{-10pt}
\subsection{Ablation Analysis}
In the above section, our proposed system has demonstrated good performance in speaker anonymization. Now, we will analyze the effectiveness of each proposed module and constraint to gain insights into their contributions to the overall performance.

Both the content extractor and speaker encoder are first analyzed. To assess their individual contributions, we replace the soft content extractor in our proposed system with the conventional phonetic posteriorgrams (PPG) extractor denoted as Variant with PPG. Instead of using the frame-level speaker vectors in our proposed speaker encoder, we extract utterance-level speaker identity representations from the pre-trained ASV model to replace the fixed-dimension speaker vector in our proposed approach, denoted as Variant with utterance. By making these changes, we can evaluate the impact of each module on the overall system's performance in speaker anonymization.

Table~\ref{tab:ablation1} shows the comparison results between the proposed system and its variants on VPC metrics. 
When we replace the soft content representation with a PPG extractor, the WER, pitch, and distinctive metrics are degraded compared to the proposed system. These results highlight the effectiveness of soft linguistic content in achieving better intelligibility and para-linguistic preservation. The soft content extractor plays a crucial role in preserving para-linguistic information during anonymization.
As for extracting utterance-level speaker identity representation, the EER results show significant degradation. This is likely because utterance-level representations may not adequately adapt to the conversion model, leading to privacy leakage in more challenging attack scenarios. The frame-level speaker vectors used in our proposed system are better suited for maintaining speaker privacy and preventing such degradation, highlighting the importance of the frame-level representations in our system's performance.

\begin{table}[ht]
\caption{Comparison results between the proposed system and its variants on VPC metrics. EER results is evaluated by $ASV_{\textrm{eval}}^{\textrm{anon}}$ in Semi-informed scenario.}\label{tab:ablation1}
\renewcommand\arraystretch{1.1}
\resizebox{1.0\linewidth}{!}{
\begin{tabular}{lcccc}
\hline
\multicolumn{1}{c}{Models} & EER $\uparrow$  & WER $\downarrow$ & $\rho^{\textrm{F0}}$ $\uparrow$    & $G_{\textrm{vd}}$ $\uparrow$     \\ \hline
Proposed                   & 39.74 & 6.43 & 0.84 & -0.98 \\
Variant with PPG             & 38.16    & 7.65   & 0.75   & -2.49    \\
Variant with utterance             & 29.17    &6.89   & 0.78   & -3.43    \\ \hline
\end{tabular}
}
\end{table}
\vspace{-10pt}

\begin{table}[ht]
\caption{Ablation results of loss constraint using VPC metrics. EER results are evaluated by $ASV_{\textrm{eval}}^{\textrm{anon}}$ in Semi-informed scenario.}\label{tab:ablation2}
\renewcommand\arraystretch{1.0}
\resizebox{1.0\linewidth}{!}{
\begin{tabular}{lcccc}
\hline
\multicolumn{1}{c}{\fontsize{6}{8}\selectfont Models} & \fontsize{6}{8}\selectfont EER $\uparrow$  & \fontsize{6}{8}\selectfont WER $\downarrow$ & \fontsize{6}{8}\selectfont $\rho^{\textrm{F0}}$ $\uparrow$    & \fontsize{6}{8}\selectfont $G_{\textrm{vd}}$ $\uparrow$     \\ \hline
\fontsize{6}{8}\selectfont Proposed                   & \fontsize{6}{8}\selectfont 39.74 & \fontsize{6}{8}\selectfont 6.43 & \fontsize{6}{8}\selectfont 0.84 & \fontsize{6}{8}\selectfont -0.98 \\
\quad \fontsize{6}{8}\selectfont -$\mathcal{L}_\textrm{mi}$             & \fontsize{6}{8}\selectfont 36.24    & \fontsize{6}{8}\selectfont 6.79   & \fontsize{6}{8}\selectfont 0.79   & \fontsize{6}{8}\selectfont -2.24    \\
\quad \fontsize{6}{8}\selectfont -$\mathcal{L}_\textrm{cls}$             & \fontsize{6}{8}\selectfont 27.76    &\fontsize{6}{8}\selectfont 6.85   & \fontsize{6}{8}\selectfont 0.79   & \fontsize{6}{8}\selectfont -1.93    \\ \hline
\end{tabular}
}
\vspace{-10pt}
\end{table}
Furthermore, we conduct ablation studies to demonstrate the effectiveness of the loss constraints in our anonymization system, including MI constraints $\mathcal{L}_\textrm{mi}$ and auxiliary classification constraints $\mathcal{L}_\textrm{cls}$ in the speaker encoder. 
The results in Table~\ref{tab:ablation2} show that when we remove the MI constraints, all metrics degrade, indicating that the MI constraints enhance the disentangling ability for the content, speaker identity, and prosody, benefiting the reconstruction of anonymized speech. Additionally, removing the auxiliary classification in the speaker encoder results in an EER of 27.76\%, demonstrating that the auxiliary classification effectively constrains the speaker-score matrix to retain speaker-related information. This highlights the importance of both MI constraints and auxiliary classification in achieving better performance in our anonymization system. 

\vspace{-5pt}
\section{Discussion}
We will further discuss the complexity and limitations of the proposed approach in real-world applications. The comparison of the complexity and requirements between the proposed approach and previous approaches is detailed in Table~\ref{tab:discussion}. Notably, our proposed approach eliminates the need for averaging other speaker information and an additional speaker pool, a crucial component in previous approaches. While the previous approach requires the selection of candidate speaker vectors from the speaker pool, resulting in complexity in the anonymization process of O(n), this is consistent with our proposed approach. Regarding model parameters, the previous approach ranges from 8M to 40M, whereas our proposed approach is about 17M, placing it in the same order of magnitude.

\begin{table}[h]
\centering
\caption{Comparing the requirements and complexity in the anonymization process of the previous anonymization approach vs. the proposed approach.} \label{tab:discussion}
\renewcommand\arraystretch{1.0}
\resizebox{1.0\linewidth}{!}{
\begin{tabular}{lcc}
\hline
                          & Conventional    & Proposed    \\ \hline
Pre-trained model         & ASR and ASV & ASR and ASV \\
Additional speaker pool   & \CheckmarkBold & \XSolidBrush            \\
Other speaker information & \CheckmarkBold & \XSolidBrush            \\
Complexity                & O(n)        & O(n)        \\
Parameters                & 8M$\sim$40M & 17M         \\ \hline
\end{tabular}
}
\end{table}

Regarding limitations, the proposed approach can only anonymize English speakers, whereas an ideal speaker anonymization system should be language-independent. Specifically, our proposed approach utilizes a language-dependent content extractor trained on the English dataset, making it unsuitable for extracting linguistic content for unseen languages. This mismatch may significantly affect utility and intelligibility in multi-lingual speaker anonymization scenarios.

\vspace{-5pt}
\section{Conclusion}
This study proposes a novel anonymization approach that employs SVD to generate the singular value transformation-assisted matrix for anonymizing original speech.
We leverage the attention mechanism within SSL to acquire a speaker-score matrix and speaker-related tokens, effectively capturing the distinct characteristics of speaker identity. To ensure robust privacy protection and distinctiveness, we utilize SVD to decompose the speaker-score matrix and apply a non-linear transformation to the resulting singular matrix to conceal the original speaker's identity. Additionally, we leverage soft content representation and MI constraints better to disentangle content, speaker identity, and prosody for improved naturalness.
Experiments on VPC datasets demonstrate the effectiveness of our approach in protecting speaker privacy under all attack scenarios while maintaining speech naturalness and distinctiveness. Compared with the VPC baseline and the first-place system, our proposed anonymization system achieves the best performance in privacy and utility across all VPC metrics. Ablation studies further confirm the effectiveness of each component in preserving privacy and containing naturalness and distinctiveness.

%
\IEEEpeerreviewmaketitle


%

\appendices




\ifCLASSOPTIONcaptionsoff
  \newpage
\fi



\bibliographystyle{IEEEtran}
%
\bibliography{mybib}
\end{document}